\documentclass[aps,prl,preprint,superscriptaddress]{revtex4}
\usepackage{graphicx}
\usepackage{color}

\begin{document}
\title{High Crystallinity and Decoupling of Graphene on a Metal: Reduced Coulomb Screening and Tunable $pn$-Junctions}
\author{S\o ren Ulstrup}
\affiliation{Department of Physics and Astronomy, Interdisciplinary Nanoscience Center (iNANO), Aarhus University, Denmark}
\author{Mie Andersen}
\affiliation{Department of Physics and Astronomy, Interdisciplinary Nanoscience Center (iNANO), Aarhus University, Denmark}
\author{Marco Bianchi}
\affiliation{Department of Physics and Astronomy, Interdisciplinary Nanoscience Center (iNANO), Aarhus University, Denmark}
\author{Lucas Barreto}
\affiliation{Department of Physics and Astronomy, Interdisciplinary Nanoscience Center (iNANO), Aarhus University, Denmark}
\author{Bj\o rk Hammer}
\affiliation{Department of Physics and Astronomy, Interdisciplinary Nanoscience Center (iNANO), Aarhus University, Denmark}
\author{Liv Hornek\ae r}
\affiliation{Department of Physics and Astronomy, Interdisciplinary Nanoscience Center (iNANO), Aarhus University, Denmark}
\author{Philip Hofmann}
\affiliation{Department of Physics and Astronomy, Interdisciplinary Nanoscience Center (iNANO), Aarhus University, Denmark}
\email{philip@phys.au.dk}

\date{\today}
 \begin{abstract}
High quality epitaxial graphene films can be applied as templates for tailoring graphene-substrate interfaces that allow for precise control of the charge carrier behavior in graphene through doping and many-body effects. By combining scanning tunneling microscopy, angle-resolved photoemission spectroscopy and density functional theory we demonstrate that oxygen intercalated epitaxial graphene on Ir(111) has high structural quality, is quasi free-standing, and shows signatures of many-body interactions. Using this system as a template, we show that tunable $pn$-junctions can be patterned by adsorption and intercalation of rubidium, and that the $n$-doped graphene regions exhibit a reduced Coulomb screening via enhanced electron-plasmon coupling. These findings are central for understanding and tailoring the properties of graphene-metal contacts e.g. for realizing quantum tunneling devices.\\

\end{abstract}

\maketitle

\section{Introduction}

The synthesis of epitaxial graphene films has been intensively explored in the past decade, where there has been remarkable advances in graphene growth processes involving either the thermal decomposition of SiC substrates\cite{Berger:2004,Emtsev:2009} or chemical vapor deposition on metal surfaces\cite{Sutter:2008,Coraux:2008,Li:2009}. These epitaxial growth methods now seem close to achieving cost-effective large-area graphene films with high crystallinity\cite{hao:2013}, which can potentially lead to high frequency electronics and photonics devices\cite{Novoselov:2012}. Furthermore, epitaxial graphene systems exhibit very large freedom in design and modification through controllable adsorption or intercalation of a number of atomic species, tailoring the electronic\cite{Bostwick:2007,Riedl:2009,Balog:2010,Lizzit:2012,schumacher:2013} and magnetic properties\cite{Varykhalov:2008,Sicot:2012,marchenko:2012} of the two-dimensional Dirac particles in graphene.   

The $\pi$-band dispersion in graphene, characterized by a Dirac cone around the Fermi level, can be affected by the supporting substrate, and tuned through the coupling with this substrate. The idea of complete substrate decoupling by intercalation, leading to so-called quasi free-standing graphene, was developed for systems, such as single-layer graphene on Ni(111) or on SiC, where hybridization of the $\pi$-states with the substrate prevents the ideal linearly dispersing low energy spectrum from being established\cite{Varykhalov:2008,Emtsev:2008,Riedl:2009}. Additionally, the graphene-substrate interface plays a crucial role for controlling the doping level in graphene, and it affects the effective Coulomb screening of the charge carriers. Studies of quasi free-standing graphene samples on SiC substrates have revealed that tuning these properties can lead to distortions of the slope along the Dirac cone and a plasmon-related reconstruction of the Dirac point has been observed \cite{Bostwick:2010,Siegel:2011}. Interestingly, such effects have only been reported for graphene in proximity to a semiconducting substrate, and are typically assumed to be completely screened for graphene grown on metal surfaces\cite{Siegel:2012}. However, the conditions in the interface to the metal might substantially modify this picture\cite{Walter:2011c}.  

Here we explore oxygen intercalated graphene on Ir(111), which was recently demonstrated as a viable route to decouple an entire single layer of graphene from its Ir(111) substrate\cite{Larciprete:2012}. In contrast to e.g. Ni(111) or SiC, where it is challenging to achieve high quality single layers of graphene because of rotational domains\cite{zhao:2011}, or nucleation of multilayers at step-edges\cite{Emtsev:2009}, graphene growth on Ir(111) self-terminates once the monolayer is complete, and single-domain samples can be routinely achieved. These samples exhibit a weakly $p$-doped Dirac cone, which shows signatures of substrate-related effects, such as hybridization gaps and replica bands\cite{Pletikosic:2009,Kralj:2011,Starodub:2011}. Such effects greatly complicate studies of the underlying electron dynamics and many-body interactions in graphene, requiring the intercalation step to recover these properties.

Using scanning tunneling microscopy (STM), low energy electron diffraction (LEED) and density functional theory (DFT) we investigate the structural properties of the intercalated system, and demonstrate that it has excellent crystalline quality, with no signs of oxygen induced damage of the carbon lattice. In angle-resolved photoemission spectroscopy (ARPES) measurements we examine the effect of intercalation on both the graphene $\pi$-band and the surrounding Ir valence states, and show that the entire $\pi$-band dispersion is unaffected by the metallic substrate. Due to these excellent properties, we employ this sample as a platform for investigating many-body effects in intercalated graphene on a metal. Through electron doping by an additional Rb intercalation step, we study the effective screening properties of the graphene-metal interface. With ARPES and DFT we demonstrate that partial Rb intercalation leads to a tunable double doping structure, realizing a tunable quasi free-standing graphene $pn$-junction on a metal contact.   

\section{Methods}

A single crystal Ir(111) sample was cleaned by sputtering with 2~keV Ne ions and flashing to 1520~K followed by cycles of annealing to 1070~K in an O$_2$ background pressure of $5\times10^{-8}$~mbar to remove carbon impurities. The remaining O$_2$ on the surface was removed by annealing to 1000~K in a H$_2$ background pressure of $5\times10^{-7}$~mbar. This procedure was repeated until a sharp hexagonal LEED pattern from the Ir(111) surface was observed. A complete monolayer of graphene was grown on the clean Ir(111) surface by doing at least 15 temperature-programmed growth cycles, where an ethylene background pressure of $5\times10^{-7}$~mbar was kept while ramping the temperature between 520~K and 1520~K. The completeness of the graphene layer was checked both with LEED, where a clear moir\'e pattern appears, and with ARPES, where a sharp Dirac cone is observed. The sample temperature was measured using a K-type thermocouple in thermal contact with the non-polished side of the Ir crystal, and from the readings of an infrared pyrometer, facing the polished side of the crystal.   

Intercalation of O$_2$ was achieved by fully encapsulating the sample surface in a custom-made doser consisting of a molybdenum nozzle attached to an O$_2$ leak valve, and keeping the sample temperature at 520~K while dosing O$_2$ at a background pressure of $5\times10^{-4}$~mbar for 10 minutes. Rubidium was intercalated by evaporating the material from a commercial getter source facing the surface of the sample, while keeping the sample at room temperature. 

The oxygen intercalated graphene samples were prepared in the ARPES UHV chamber, and then transferred \emph{ex situ} to the STM chamber, where contaminants were desorbed by heating to 400~K. STM measurements on non-intercalated graphene on Ir(111) were done on \emph{in situ} prepared samples. In all chambers the base pressures were better than $2\times10^{-10}$~mbar during measurements.

STM experiments were conducted using an Aarhus-type instrument\cite{laegsgaard:1988}. ARPES and LEED measurements were done at the SGM3 beamline of the ASTRID synchrotron light source in Aarhus, Denmark\cite{Hoffmann:2004}. Electronic band structure maps were acquired using a photon energy of 100~eV with total energy- and angular-resolution of 26~meV and 0.2$^{\circ}$, respectively, while detailed scans around the Dirac point were taken with a photon energy of 47~eV and total energy- and angular-resolution of 16~meV and 0.1$^{\circ}$, respectively. The sample temperature was kept at 70~K using a closed-cycle He cryostat during ARPES and LEED measurements.\\  

The DFT calculations were performed with the real-space projector augmented wave code, GPAW\cite{Enkovaara2010} and the ASE interface\cite{Bahn2002}. Since the investigated systems contain dispersive interactions between the graphene sheet and its substrate, a functional which takes into account vdW interactions was chosen, specifically the optB88-vdW functional \cite{Klimes2010}, which has proven successful for similar systems such as graphene on Ni \cite{Mittendorfer2011} and benzene adsorption on metals \cite{Liu2012}. 

Two different approximations for the graphene moir\'e superstructure was used as described in the text, namely a ($9\times9$) graphene to an ($8\times8$) Ir slab and a rotated ($6\times6$) graphene to a ($\sqrt{28}\times\sqrt{28}$) Ir slab. For all structures three Ir layers were used with the bottom Ir layer kept fixed, while all other atoms were relaxed until the maximum force on each atom was below 0.02 eV/\AA. The graphene lattice constant was fixed to its optimized value of 2.465~\AA, and the Ir lattice constant was adapted accordingly, resulting in a strain of about 1~\% for both cells. 2D periodic boundary conditions were employed parallel to the surface, and a vacuum region of 6~\AA~ separated the slabs from the cell boundaries perpendicular to the surface. A $(2\times2)$ \textit{k}-point sampling was used for all structural optimizations and energies, whereas the density of states calculations were performed with a $(6\times6)$ \textit{k}-point grid for the small cell and a $(4\times4)$ \textit{k}-point grid for the large cell. The grid spacing was 0.185~\AA.

Rb adsorption potential energies were calculated as $(E_{\mathrm{GR/Rb/O/Ir}}-E_{\mathrm{GR/O/Ir}}-n\cdot E_{\mathrm{Rb}})/n$, where $E_{\mathrm{GR/Rb/O/Ir}}$, $E_{\mathrm{GR/O/Ir}}$ and $E_{\mathrm{Rb}}$ are the energies of the mixed O and Rb intercalated structure, the O intercalation structure and a Rb atom, respectively, and where $n$ is the number of Rb atoms in the unit cell. Graphene binding energies were calculated as $(E_{\mathrm{X/Ir}}+E_{\mathrm{GR}}-E_{\mathrm{GR/X/Ir}})/m$, where $E_{\mathrm{X/Ir}}$, $E_{\mathrm{GR}}$ and $E_{\mathrm{GR/X/Ir}}$ are the energies of the Ir surface with adsorbates X, the isolated graphene and GR/Ir intercalated with X, respectively, and where $m$ is the number of C atoms in the graphene. In this work X replaces either O atoms, Rb and O atoms or nothing (non-intercalated GR/Ir).

\section{Results and Discussion}

An STM investigation of the structure of oxygen intercalated graphene on Ir(111) (GR/O/Ir) is presented in Fig. 1(a)-(c). The graphene layer extends continuously over the step-edges as seen in Fig. 1(b), proving that these are not reactive under the intercalation conditions. In Fig. 1(c) a moir\'e superstructure with lattice constant 25.3~\AA~as in plain graphene on Ir(111) (GR/Ir) can be observed. Since the structural coherency of the graphene film is found over very large areas as demonstrated in Fig. 1(a), the graphene layer is homogeneously oxygen intercalated with an intact, defect-free carbon lattice. The intercalation of such large graphene sheets is believed to occur via the microscopic wrinkles appearing in the sheet during the graphene growth process\cite{petrovic:2013}. Once the oxygen atoms penetrate into the graphene-Ir interface
they diffuse and bind to the Ir substrate, leaving an intact carbon lattice above\cite{Larciprete:2012}.

\begin{figure} [t!]
\includegraphics[width=.45\textwidth]{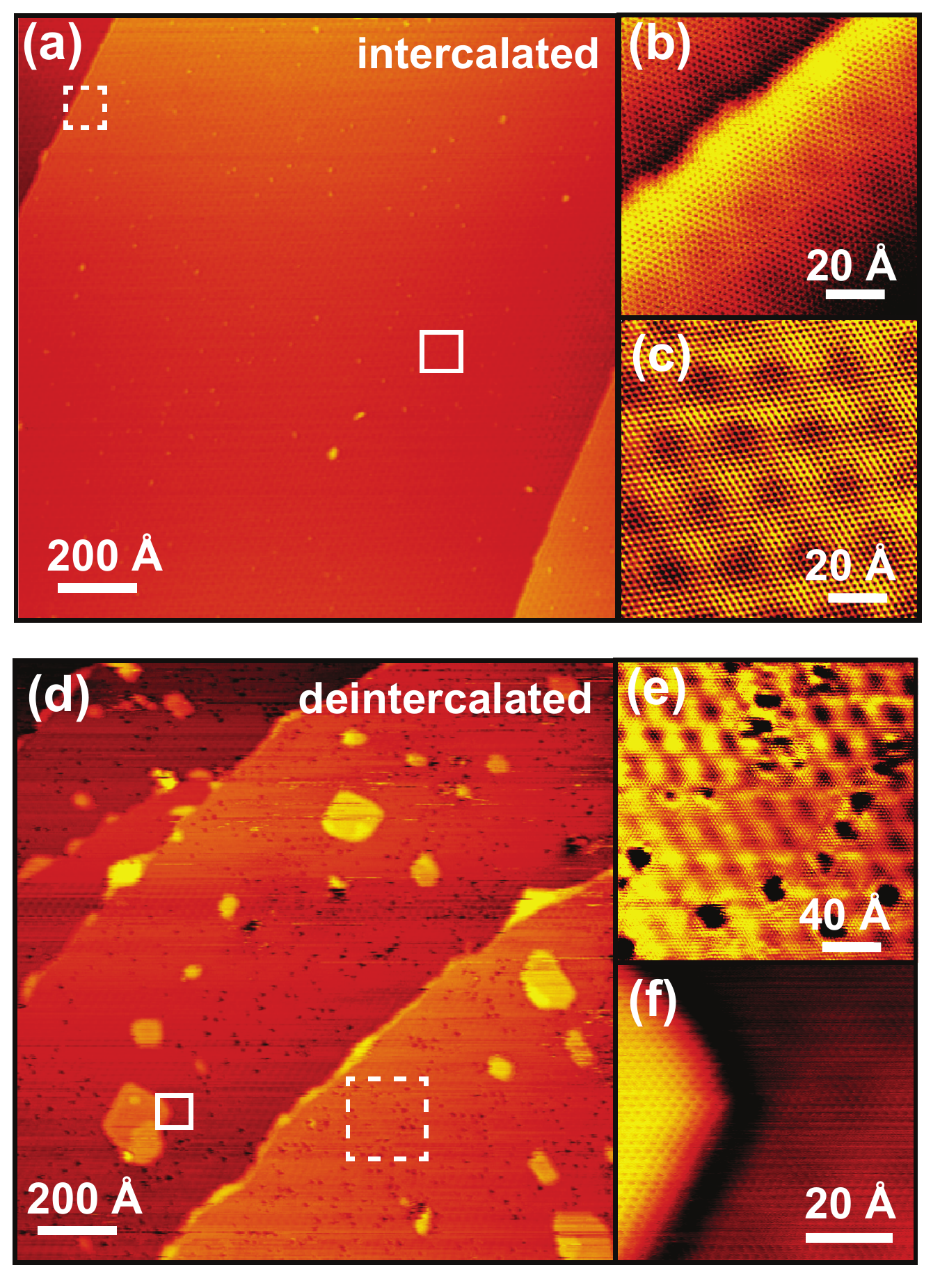}\\
\caption{Structural quality and stability of oxygen intercalated graphene on Ir(111): (a)-(c) STM images of intercalated graphene extending continuously over terraces and step-edges. The large-area image in (a) was taken with tunneling parameters $I_t=0.930$ nA and $V_t = 20.8$ mV. The atomic resolution images of the (b) step-edge region marked by the dashed square in (a), and the (c) graphene lattice within the full square in (a) were obtained with tunneling parameters $I_t=1.100$ nA and $V_t = 2.1$ mV. (d)-(f) STM images after oxygen deintercalation by heating above 600~K. (e) Image corresponding to the area marked by the dashed square in (d) showing large etch holes in the graphene lattice. (f) Zoom-in of the edge of the blister marked by the full square in (d). The graphene lattice is seen to extend over this structure. In (d)-(f) the tunneling parameters were fixed at $I_t=1.100$ nA and $V_t = 2.1$ mV.}
\label{fig:1}
\end{figure}

The intercalated system is stable up to a temperature of 600~K. Heating above this point yields the deintercalated graphene seen in Fig. 1(d)-(f). Etch holes and blisters with apparent heights in the range of 1-2~\AA~are observed both in the middle of terraces and along step-edges. In Fig. 1(e) it is seen that etch holes occur predominantly in the bright regions of the graphene moir\'e, consistent with these regions being reactive as observed for hydrogen adsorption on GR/Ir\cite{Balog:2010}. From these experiments it generally appears that the etching does not occur in a periodic fashion on this substrate. In Fig. 1(f) the edge of a blister is shown. The graphene lattice extends across the edge of this structure, which might be a graphene bubble caused by remaining oxygen. 

A structural comparison between newly prepared GR/Ir and GR/O/Ir is presented in Fig. 2. In the atom-resolved STM images in Fig. 2(a) and (b) the bright, dim and dark regions of the moir\'e are much less pronounced in the intercalated sample. This is studied quantitatively in Fig. 2(c) showing line profiles of the apparent height modulation $h$ through these regions for the two samples. The fast variation in these curves is due to the carbon lattice, while the underlying slow variation is the moir\'e structure. The latter is extracted by fitting polynomials to the profiles as shown in Fig. 2(c). The distance between minimum and maximum of these polynomials is taken as a measure of the graphene corrugation $\Delta d$. Admittedly such a corrugation analysis is complicated by the facts that the STM image contrast depends on the tunneling parameters and the state of the tip\cite{borca:2010,Diaye:2008}. However, we proceeded by analyzing more than 50 images like those in Fig. 2(a) and (b) of both samples, all taken with the same tunneling parameters but for different spots on the sample and different sample preparations. This yields an estimate of $\Delta d_{\mathrm{GR/Ir}} = 1.09 \pm 0.25$ \AA~and of $\Delta d_{\mathrm{GR/O/Ir}} = 0.26 \pm 0.09$ \AA. Based on these values the oxygen-intercalated graphene is expected to be roughly a factor of four less corrugated than graphene on Ir(111).

\begin{figure*} [t!]
\includegraphics[width=.85\textwidth]{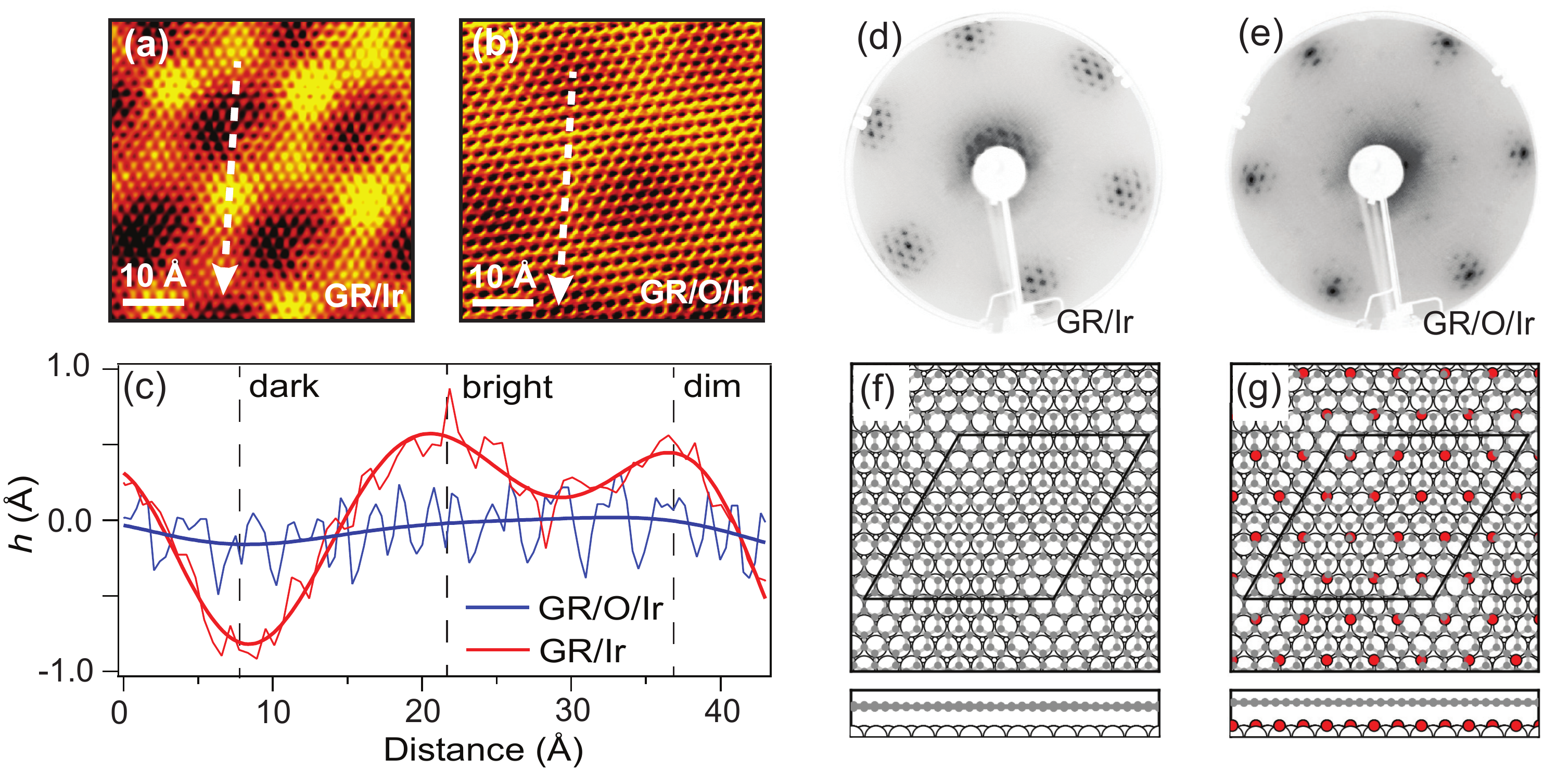}\\
\caption{Tuning the graphene superstructure by intercalation: (a)-(b) Atomic resolution STM images of GR/Ir and GR/O/Ir obtained with tunneling parameters $I_t=1.100$ nA and $V_t = 2.1$ mV. (c) Apparent height modulation $h$ along the dashed lines in (a) and (b) shown by thin lines, while thick lines are polynomial fits to extract the slow variation arising from the moir\'e lattice. (d)-(e) LEED images of (d) GR/Ir and (e) GR/O/Ir taken with a kinetic energy of 140~eV. (f)-(g) Relaxed configurations (top view and side view) of (f) GR/Ir and (g) GR/O/Ir displaying the ($2\times2$) structure of O (red circles) with respect to Ir (white circles). Carbon atoms are marked with grey circles and unit cells are outlined by black lines.}
  \label{fig:2}
\end{figure*}

LEED images of the two samples are shown in Fig. 2(d) and (e). A faint $(2\times2)$ pattern is seen relative to the hexagonal Ir(111) pattern in Fig. 2(e), which stems from the $(2\times2)$ adsorbate structure that the intercalated oxygen forms on the Ir surface. Comparing the LEED patterns of GR/Ir and GR/O/Ir it is evident that the higher order moir\'e spots visible in GR/Ir diminish upon oxygen intercalation in good agreement with the observation of a less corrugated surface in STM. Furthermore, the six principal graphene related spots gain intensity and become distinguishable from the Ir spots and moir\'e satellites, signifying that the graphene is now less interacting with the underlying surface. 

The GR/Ir and GR/O/Ir structures are investigated by DFT calculations as shown in Fig. \ref{fig:2}(f)-(g). All calculations throughout this work have been carried out using a functional that takes the van der Waals interactions in the system into account (see methods for details). In order to accommodate the intercalated ($2\times2$) structure of O with respect to Ir in the unit cell, the GR/Ir moir\'e superstructure is described by a ($9\times9$) graphene layer resting on top of a ($8\times8$) Ir slab (see Fig. \ref{fig:2}(f)-(g)), which is a fair approximation to the experimentally observed incommensurate structure of a ($10.32\times10.32$) graphene layer on a ($9.32\times9.32$) Ir slab \cite{Diaye:2008}. The most stable position of the O atoms in the intercalated structure is found to be the Ir fcc hollow sites, which is also the most stable position on the clean Ir(111) surface \cite{Krekelberg_JPCB_2004}. The average graphene binding energy (measured per carbon atom) decreases from 63~meV in GR/Ir to 55~meV in GR/O/Ir corresponding to a weakening of 8~meV in the intercalated case. The average Ir-graphene distance increases from 3.53~\AA~in GR/Ir to 4.02~\AA~in GR/O/Ir, and a reduction by nearly a factor of 3 in the graphene corrugation from 0.38~\AA~to 0.14~\AA~is found upon intercalation. The reduced corrugation is in line with the observations in the experiments. All results are summarized in Tab. 1.
 
The ARPES data presented in Fig. 3 provide an overview of the electronic structure for both GR/Ir and GR/O/Ir within a segment of the graphene and Ir Brillouin zone (BZ). The constant energy map at 0.45~eV in Fig. 3(a) is characterized by small circular energy surfaces around $\bar{K}$, which are the Dirac cone and its replicas. The asymmetric intensity distribution around the circular energy contour is a well-known interference effect in graphene\cite{Shirley:1995b,mucha:2008}. Ir-related features otherwise dominate the BZ. The $\pi$-band dispersion for GR/Ir with weakly $p$-doped Dirac cones at the $\bar{K}$-points and the saddle point at $\bar{M}$ are seen in Fig. 3(c). In addition, hybridization gaps around 2~eV caused by Ir $d$-levels crossing the $\pi$-band and minigaps around 1~eV at the mini BZ boundaries induced by the moir\'e pattern appear in the circled parts of the band in Fig. 3(c)\cite{Pletikosic:2009}. 

\begin{figure} [t!]
\includegraphics[width=.49\textwidth]{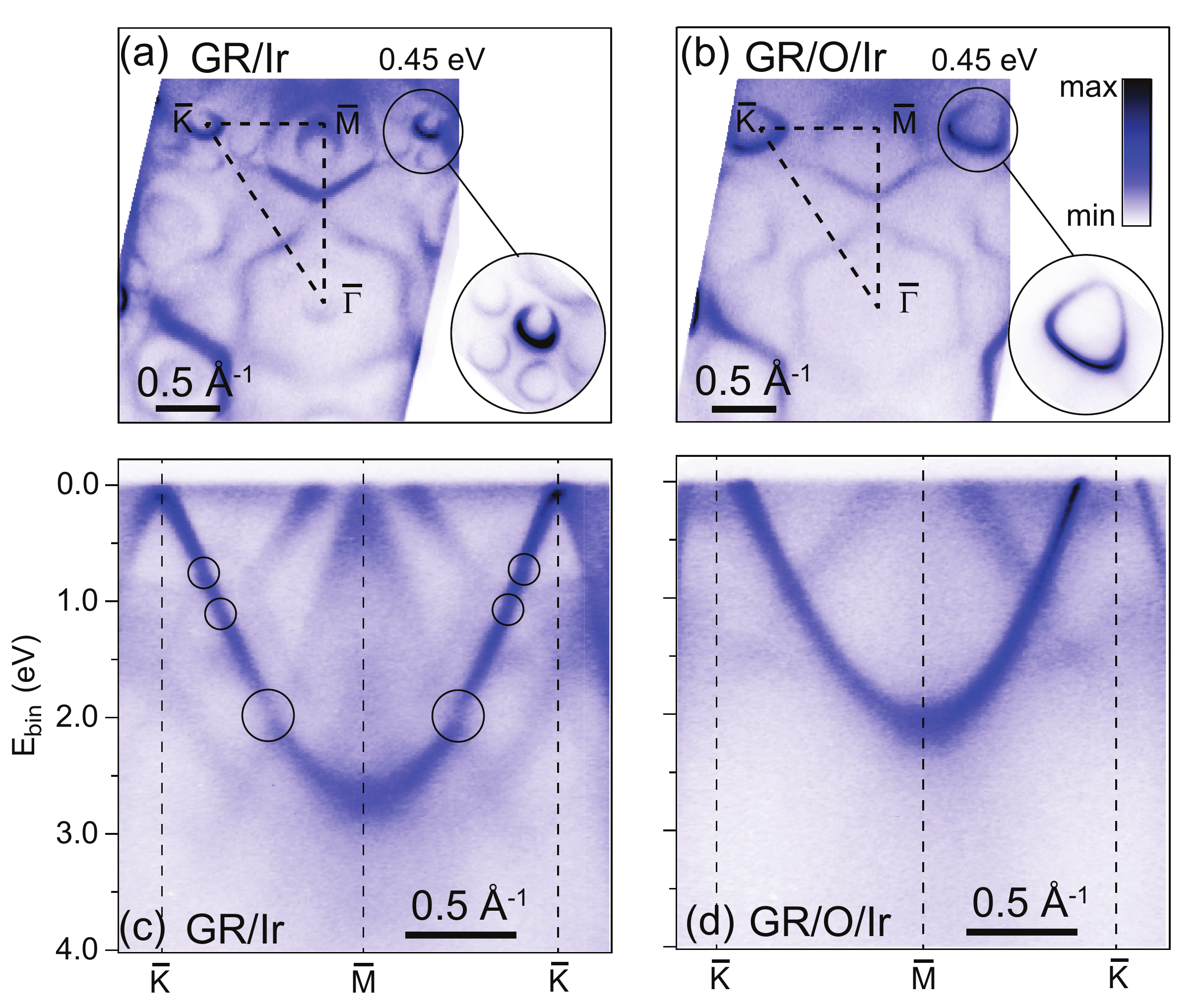}\\
\caption{Electronic structure overview of GR/Ir and GR/O/Ir: (a)-(b) Constant energy surface at a binding energy of 0.45 eV of (a) GR/Ir, and (b) GR/O/Ir. The dashed triangles mark the high-symmetry routes in the Brillouin zone of graphene. The insets provide high-resolution ARPES data of the region of $k$-space around the Dirac point. (c)-(d) Cuts along the $\bar{K}-\bar{M}-\bar{K}$ direction for (c) GR/Ir, and (d) GR/O/Ir. The circles in (c) mark substrate related gaps in the $\pi$-band of graphene.}
  \label{fig:3}
\end{figure}

Following intercalation we find as previously\cite{Larciprete:2012} a doping related shift of the entire $\pi$-band and that the Dirac cone replicas have disappeared in the energy contour of GR/O/Ir in Fig. 3(b) consistent with the smaller structural corrugation found with STM. The substrate-related gaps are no longer present in the dispersion, as shown in Fig. 3(d), confirming that the oxygen intercalated graphene is indeed completely decoupled from the substrate i.e. quasi free-standing.

Several Ir-related features appear to be disrupted once O is intercalated such as the faint circular contour at the $\bar{\Gamma}$-point in Fig. 3(a), which is a hole-like Ir surface state\cite{varykhalov:2012r}. The ellipsoidal contour around the $\bar{M}$-point in Fig. 3(a), which corresponds to the band that disperses up to the Fermi level at $\bar{M}$ in Fig. 3(c) has also vanished in Fig. 3(b) and (d). This would be consistent with this band being a surface state that is quenched by the adsorbed oxygen. Note that bulk Ir features still appear in the oxygen intercalated sample. Thus, while the graphene dispersion has been decoupled there is still a background of metallic states in other regions of the BZ, which is expected to play a large role if this sample was applied in relation to transport or optical experiments.

High resolution ARPES data of the Dirac cone cut in the $\bar{\Gamma}-\bar{K}$ direction as well as in the orthogonal direction are presented and analyzed for three different doping regimes of graphene in Fig. 4. For GR/Ir in Fig. 4(a) and (d) we see the replicas and the minigaps, and a small band gap can be discerned due to hybridization with an Ir surface state around the Fermi level. Again for GR/O/Ir in Fig. 4(b) and (e) the dispersion appears sharp and quasi free-standing. By extrapolation of the linear bands in Fig. 4(d) and (e) we can determine the location of the Dirac point to change from a binding energy of -0.07~eV in GR/Ir to -0.80~eV in GR/O/Ir corresponding to a substantial $p$-doping of $4.7\times10^{13}$~cm$^{-2}$. This value is higher than previously reported for this type of sample\cite{Larciprete:2012}, possibly because the oxygen intercalation is more complete here. To counterdope we deposit Rb atoms, and find that at the saturation level these intercalate in the graphene-Ir interface giving rise to mixed Rb and O intercalated graphene on Ir(111) (GR/Rb/O/Ir). We will discuss the case of lower Rb coverages later. For now this system allows us to study the Dirac point region, since this shifts to a binding energy around 1.33~eV as seen in Fig. 4(c) and (f), corresponding to an $n$-doping of $1.2\times10^{14}$~cm$^{-2}$.

\begin{figure*} [t!]
\includegraphics[width=.95\textwidth]{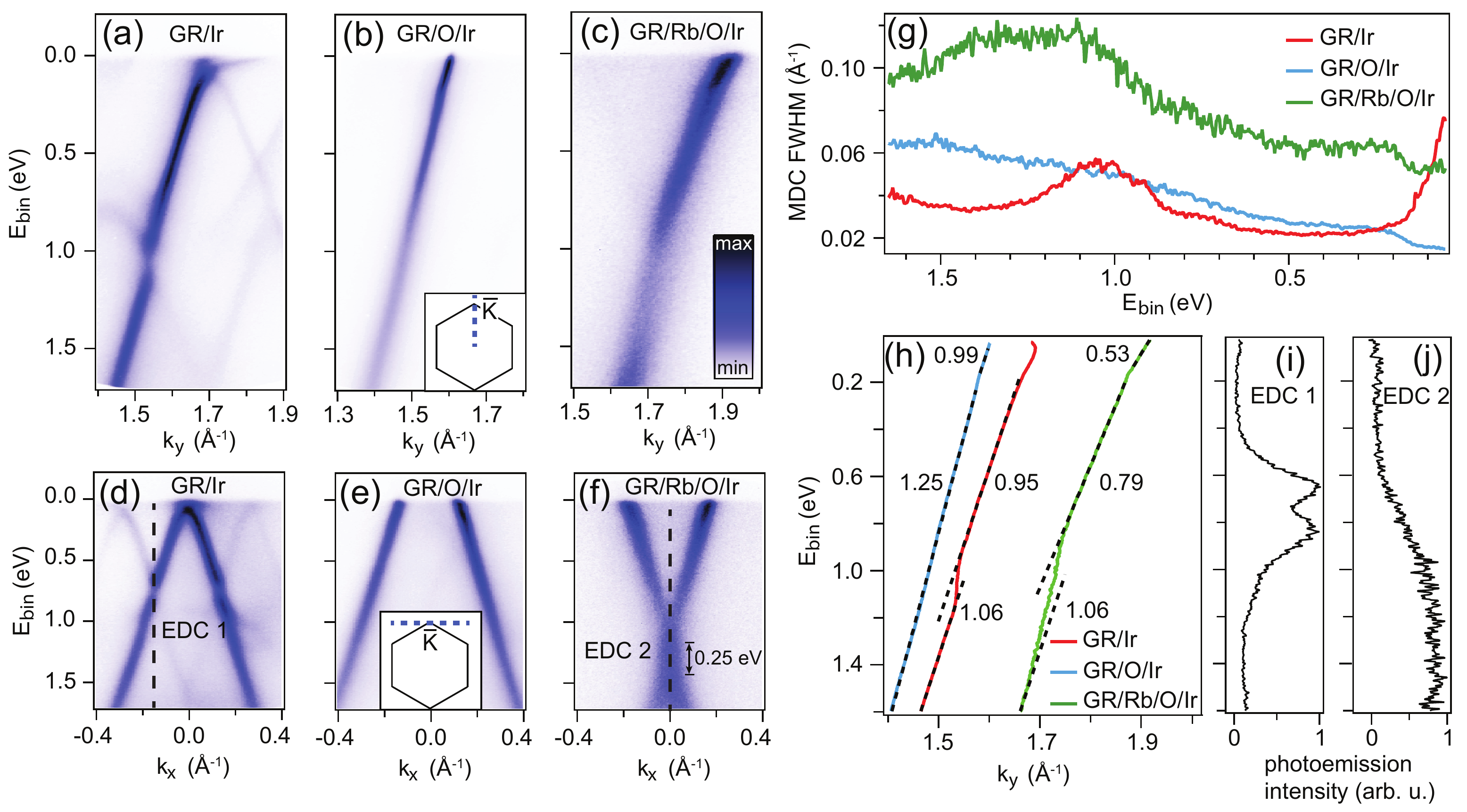}\\
\caption{Measurements and analysis of the Dirac cone, tuned by O and Rb intercalation: (a)-(c) Dispersion along the $\bar{\Gamma}-\bar{K}$ direction as shown in the inset of (b) and (d)-(f) perpendicular to this direction as shown in the inset of (e) for (a,d) GR/Ir, (b,e) GR/O/Ir, and (c,f) GR/Rb/O/Ir. In (f) an offset of 0.25~eV between valence and conduction bands is marked by a double headed arrow. (g) Full width at half maximum (FWHM) and (h) peak positions of single Lorentzian fits to MDCs of the data in (a)-(c). The dashed lines in (h) are linear fits to segments of the dispersion and the associated numbers are the band velocities given as $v^{\ast}/v$, where $v = 10^6$~m/s. (i)-(j) EDCs taken along the dashed lines in (d) and (f).}
  \label{fig:4}
\end{figure*}

The ARPES data in Fig. 4(a)-(f) can be interpreted in terms of a spectral function\cite{Hofmann:2009b} $A(\omega,\mathbf{k}) = \pi^{-1}|\Sigma^{\prime\prime}(\omega)|/\left[(\hbar\omega-\epsilon(\mathbf{k})-\Sigma^{\prime}(\omega))^{2} + \Sigma^{\prime\prime}(\omega)^2\right]$ such that the renormalization of the bare band $\epsilon(\mathbf{k})$ and the linewidth are due to the real and imaginary parts of the electronic self-energy $\Sigma^{\prime}(\omega)$ and $\Sigma^{\prime\prime}(\omega)$, respectively. Assuming that the self-energy has no strong dependence on momentum we can directly relate the linewidth procured from Lorentzian fits to momentum distribution curves (MDCs) through the data at each binding energy to $\Sigma^{\prime\prime}(\omega)$, and thus the electronic scattering mechanisms in the system\cite{Hofmann:2009b}. The MDC linewidths and renormalized dispersions extracted from the data in Fig. 4(a)-(c) are presented in Fig. 4(g) and (h), respectively. Starting with the GR/O/Ir system the linewidth close to the Fermi level is comparable to the momentum-resolution limit of the experiment, signifying a very low level of impurity scattering. Towards higher binding energies many-body effects contribute a higher linewidth. The step-like increase at 0.17~eV is caused by the excitation of an optical phonon mode, which also causes a kink in the dispersion as seen in Fig. 4(h). The further increase in linewidth is caused by electron-hole pair generation, consistent with a Fermi liquid type of behavior\cite{Das-Sarma:2007,Bostwick:2007}. For GR/Ir it is not possible to discern these effects due to the stronger coupling with the substrate. The large linewidth around the Fermi level and the hump around 1~eV seen in Fig. 4(g) are caused by the Ir surface state hybridization and minigap, respectively. These gaps also cause the distortions of the dispersion seen in Fig. 4(h). The entire linewidth of the $n$-doped GR/Rb/O/Ir sample is offset by 0.04~\AA$^{-1}$ compared to the other samples. This rigid increase in linewidth can be attributed to electron-impurity scattering induced by the deposited Rb atoms. Additionally, the electron-phonon and electron-hole processes seen in GR/O/Ir also occur in this sample, and a hump around the Dirac point region can be observed, which we will discuss shortly.

The variation between the values of the renormalized band velocities $v^{\ast}$ relative to $v=10^6$~m/s in the three systems in Fig. 4(h) are caused by a number of effects such as the electron-phonon and electron-electron interactions mentioned above\cite{DasSarma:2013}. Subtle changes in the single-particle (bare band) dispersion may also contribute. For GR/Ir the slope is not defined at the Fermi level due to the Ir-related hybridization gap. The different Fermi velocities for GR/O/Ir and GR/Rb/O/Ir can be attributed to a stronger electron-phonon induced kink in the dispersion at 0.17~eV, which is due to an enhancement of the electron-phonon coupling strength $\lambda$ following the increase in carrier concentration and thereby larger phase space available for scattering processes. This is also what we find quantitatively by extracting $\lambda$ for the dispersions in Fig. 4(a) and 4(c) using the method developed in Ref. \cite{Pletikosic:2012}. We obtain $\lambda = 0.06(9)$ for GR/O/Ir and $\lambda = 0.11(4)$ for GR/Rb/O/Ir. The slightly larger $\lambda$ value for GR/O/Ir than reported in Ref. \cite{Johannsen:2013} is consistent with the larger $p$-doping achieved for the sample in this study.      

By extrapolating the bands above and below the Dirac point in Fig. 4(f) we find that the conduction band minimum is located at a binding energy of 1.19~eV, while the valence band maximum is found at 1.44~eV. It is not possible to describe this separation of 0.25~eV between the bands in terms of a gap since the photoemission intensity peaks throughout this energy range as seen in the energy distribution curve (EDC) through the Dirac point in Fig. 4(j). This is more clear when comparing this to the EDC through the 0.15~eV minigap in GR/Ir shown in Fig. 4(i), where the gap is defined through a depression of the photoemission intensity. Broadening of the EDC arising from differently doped areas can be ruled out based on the fact that away from the Dirac point the EDCs are much more narrow. The elongated shape of the Dirac point and the hump in the MDC linewidth in Fig. 4(g) therefore seem to be better explained by a many-body effect arising from electron-plasmon coupling consistent with the situation in graphene on SiC with a carbon buffer layer in the interface (GR/SiC)\cite{Bostwick:2007}. Following the methodology of Ref. \cite{Walter:2011c} we can apply the doping level and the separation between valence and conduction bands to directly estimate the value of the effective coupling constant $\alpha = e^2/4\pi\epsilon_0\epsilon\hbar v$, where $\epsilon = (1+\epsilon_S)/2$ is the effective dielectric constant and $\epsilon_S$ is its contribution from the substrate. We find $\alpha = 0.1$ corresponding to $\epsilon = 22$ and $\epsilon_S = 43$.  This value was also found for GR/SiC\cite{Walter:2011c}. The similarity is surprising given the presence of the bulk metal in our case compared to the bulk semiconducting SiC substrate.

We will now focus on lower Rb coverages. Interestingly, we observe the emergence of two Dirac cones giving rise to $p$-type and $n$-type Fermi surfaces and dispersions as seen in Fig. 5(a)-(h). This effect is caused by the build-up of two phases that we denote as phase 1 and phase 2. We assign phase 1 to Rb atoms adsorbing on top of graphene in the GR/O/Ir system, while in phase 2 we propose that they intercalate giving rise to the GR/Rb/O/Ir system explored in Fig. 4 at saturation. The two phases coexist at intermediate Rb coverage, and since the ARPES measurement averages laterally over the sample, both phases contribute to the photoemission signal. Similar behavior was recently observed for Cs intercalation of GR/Ir\cite{petrovic:2013}. Here we find that phase 1 becomes progressively less $p$-doped, while phase 2 appears with a nearly fixed $n$-doping until saturation where phase 1 disappears, and phase 2 becomes slightly more $n$-doped. The carrier concentration and Dirac point position for the different phases are tracked in Fig. 5(i). Note, that at intermediate doping a reconstruction of the Dirac point can be seen in phase 2 e.g. in Fig. 5(f) consistent with the formation of a plasmaron\cite{Bostwick:2010} and the finite screening discussed above.

\begin{figure} [t!]
\includegraphics[width=.6\textwidth]{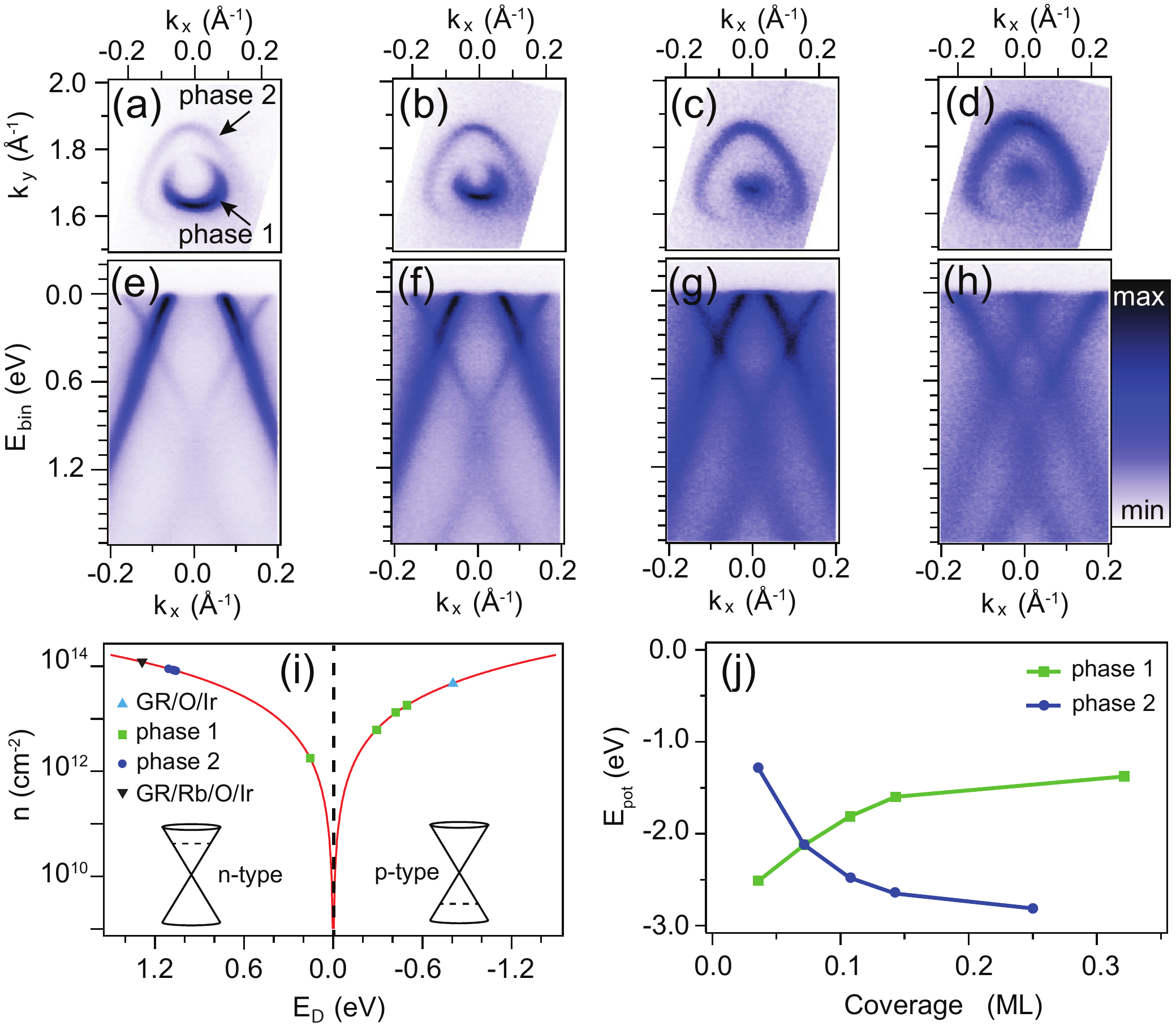}\\
\caption{Build-up of $pn$-junction structure by partial Rb intercalation: (a)-(d) Fermi surfaces, and (e)-(h) dispersions orthogonal to the $\bar{\Gamma}-\bar{K}$ direction after four consecutive Rb doses. (i) Chart of observed Dirac point energies and corresponding carrier concentrations, covering the transition from $p$-doped GR/O/Ir to $n$-doped GR/Rb/O/Ir via the two separate doping phases marked on the Fermi surfaces in (a). (j) DFT calculations of adsorption potential energies, E$_{pot}$, of adsorbed (phase 1) and intercalated (phase 2) Rb atoms as a function of Rb coverage with respect to Ir.}
  \label{fig:5}
\end{figure}

In order to corroborate our assignment of the two phases to adsorbed and intercalated Rb, we explore the energetics and structures of these phases further using the DFT calculations presented in Fig. 5(j) and Fig. 6, respectively. Due to the presence of Rb atoms making the calculations more demanding, a smaller approximative unit cell consisting of a rotated ($6\times6$) graphene layer on a ($\sqrt{28}\times\sqrt{28}$) Ir slab (see Fig. 6(a)), compared to the one in Fig. 2(f)-(g), is used. In Tab. \ref{Tab:M1} the average graphene binding energy, \textit{E$_{\mathrm{b}}$}, the average Ir-graphene distance, \textit{d$_{av}$}, and the graphene corrugation, \textit{$\Delta$d}, are compared for the two cells. The average values of graphene binding energies and distances are insensitive to the chosen unit cell, whereas a correct description of the graphene corrugation requires the larger unit cell. 

\begin{figure} [t!]
\includegraphics[width=.6\textwidth]{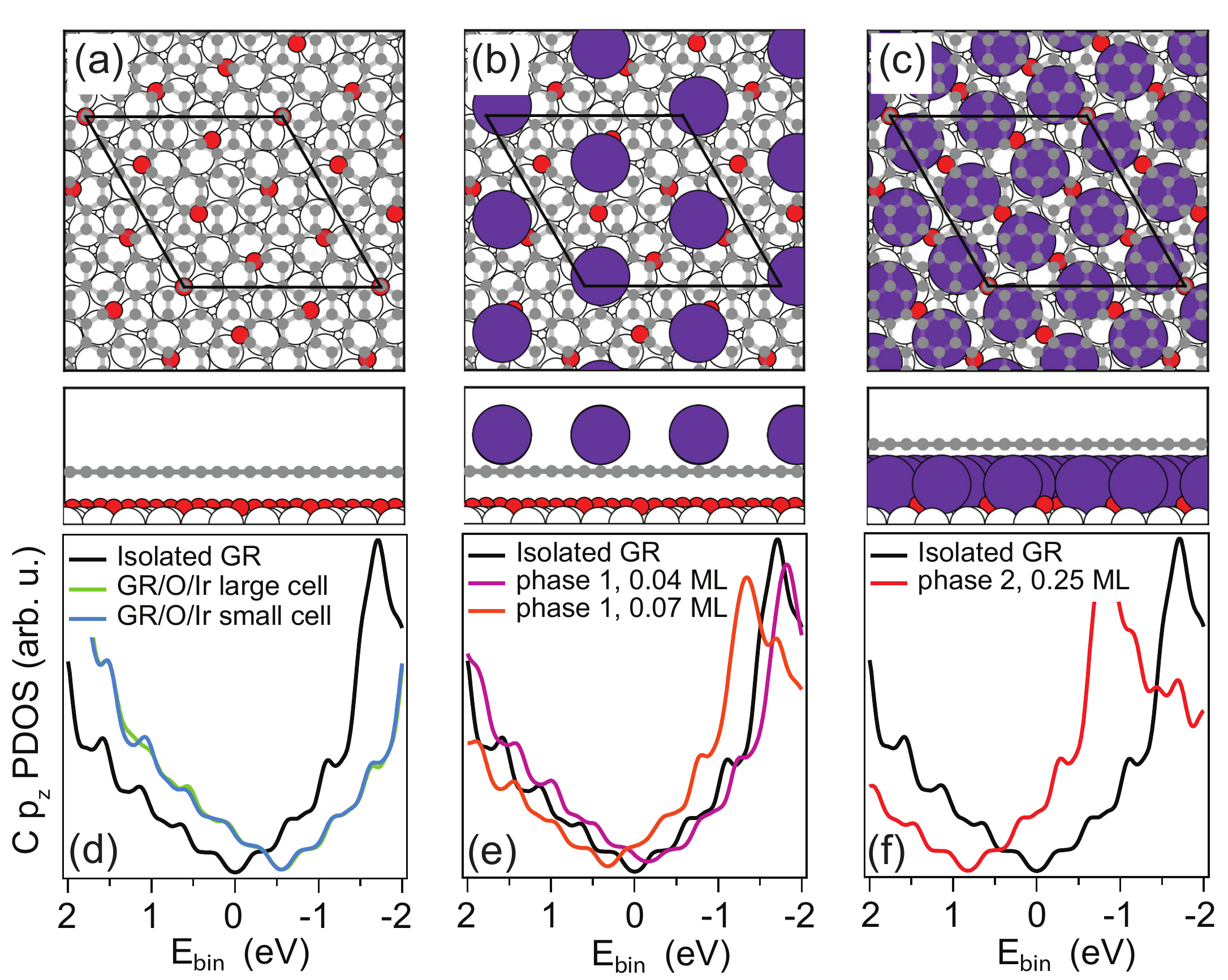}\\
\caption{Calculated doping structures: Relaxed configurations (top view and side view) and density of states projected onto the C $p_z$ orbitals for (a,d) GR/O/Ir, (b,e) Rb adatoms (purple circles) on GR/O/Ir, and (c,f) GR/Rb/O/Ir. The configuration depicted in (b) corresponds to a Rb coverage of 0.11 ML with respect to Ir (white circles). Oxygen (carbon) atoms are marked with red (grey) circles and unit cells are marked in black.}
  \label{fig:6}
\end{figure}

The three investigated structures of GR/O/Ir, Rb adatoms on GR/O/Ir (phase 1) and GR/Rb/O/Ir (phase 2) are presented in Fig. \ref{fig:6}(a)-(c). For the Rb adatom structure in Fig. \ref{fig:6}(b) it is found that the Rb atoms prefer to adsorb in the center of the carbon hexagons in agreement with previous studies\cite{Rytkonen_PRB_2007}. The position of the Rb atoms with respect to the underlying O/Ir lattice is not important for the binding energies. It is found that the Rb atoms repel each other, so that the most stable configuration for each coverage investigated corresponds to the one with the maximal distance between the Rb atoms. For the Rb intercalated structure in Fig. \ref{fig:6}(c) the Rb atoms prefer to adsorb in between two O atoms close to an Ir fcc hollow site, but slightly displaced towards the Ir top site. 

In Fig. \ref{fig:5}(j) the adsorption potential energies, E$_{pot}$, of the Rb atoms in the two phases are plotted for a coverage ranging from 0.04 ML to 0.32 ML with respect to Ir. The most stable configurations of the lower coverage Rb intercalated structures were found using a genetic algorithm approach\cite{Vilhelmsen_PRL_2012}, where the positions of the Rb atoms are varied while the O atoms stay at the fcc hollow sites. The resulting structures (not shown) all have Rb atoms adsorbed in-between two O atoms, and furthermore the distances between the Rb atoms are maximized. Comparing the energies of the two phases, it is seen that the adatom phase is preferred at low coverage up to around 0.07 ML, whereas the intercalated phase is preferred at higher coverage. This is in good agreement with the experimental results. The overall most favorable structure within the investigated range of coverage is the 0.25 ML intercalated structure. Therefore this structure should be the only intercalation structure formed in good agreement with the observed fixed doping level of phase 2 in Fig. 5. 

As discussed in Ref. \cite{petrovic:2013} the preference for dense intercalated structures derives from the loss in graphene binding energy when the graphene is lifted away from the Ir substrate in the intercalation process. With denser intercalated structures the energy penalty for lifting the graphene is divided out on more atoms, leading to a higher stability per intercalated atom. On top of this effect we furthermore find that the graphene binding energy itself is reduced less for dense intercalated structures, since the denser array of intercalated atoms leads to a more favorable van der Waals interaction between graphene and intercalants. This is shown in Tab. \ref{Tab:M1}, where we compare the average graphene binding energy for the mixed Rb/O-intercalated structure at the highest and the lowest coverage. It is seen that the graphene binding energy is reduced to 22 meV per C atom for the low coverage intercalation structure, whereas for the high coverage structure the graphene binding is 51 meV per C atom, which is similar to the value for the O-intercalated structure.
\begin{center}
  \begin{table}[t!]
    \begin{tabular*}{0.48\textwidth}{ @{\extracolsep{\fill}}l| c  c  c  c}
       & \textit{E$_{\mathrm{b}}$} & \textit{d$_{av}$} & \textit{$\Delta$d} \\\hline
GR/Ir & 61; \textit{63} & 3.50; \textit{3.53} & 0.04; \textit{0.38} \\
GR/O/Ir & 54; \textit{55} & 3.94; \textit{4.02} & 0.04; \textit{0.14} \\       
phase 2, 0.04 ML & 22 & 5.02 & 0.70 \\
phase 2, 0.25 ML & 51 & 6.10 & 0.06 \\
    \end{tabular*}
    \caption{Average graphene binding energies \textit{E$_{\mathrm{b}}$} (in meV per carbon atom), average Ir-graphene distances \textit{d$_{av}$} (in \AA), and graphene corrugations \textit{$\Delta$d} (in \AA) for GR/Ir, GR/O/Ir, GR/Rb/O/Ir (phase 2) at low and high coverages. For results given in italic the large unit cell shown in Fig. 2(f)-(g) has been used, whereas for the remaining results the smaller approximative unit cell shown in Fig. 6(a)-(c) has been used.}
  \label{Tab:M1}
\end{table}
\end{center}
We find that the binding energy of an adsorbed Rb atom on GR/Rb/O/Ir, i.e. on already Rb intercalated areas is merely 1.50~eV at the highest coverage of intercalated Rb, compared to 2.52~eV on GR/O/Ir, i.e. areas without intercalated Rb. This explains the tendency of the system to separate into adsorbed and intercalated phases and not mixed phases where Rb is both adsorbed and intercalated. 

In Fig. \ref{fig:6}(d)-(f) the density of states projected onto the C $p_z$ orbitals (PDOS) is plotted for each of the structures along with a reference PDOS corresponding to isolated graphene. The narrow peaks in these curves are an artifact of the discrete sampling of \textit{k}-points in the BZ. For GR/O/Ir (Fig. \ref{fig:6}(d)) the PDOS is plotted both for the smaller and larger unit cell, demonstrating that the PDOS is insensitive to the choice of cell. For GR/O/Ir a $p$-doping of around 0.6 eV is calculated, which is qualitatively in line with the ARPES measurements in this work. Adsorbing Rb atoms on top of the graphene sheet (Fig. \ref{fig:6}(e)) shifts the Dirac point down in energy, matching the behavior of phase 1 in the experiment. The precise value of the shift depends on the Rb coverage taking it from $p$- to $n$-doping with increasing coverage. For the mixed GR/Rb/O/Ir we only consider the PDOS for the 0.25 ML case (Fig. \ref{fig:6}(f)), since this is the energetically most favorable structure. An $n$-doping of around 0.8~eV is calculated for this structure. This is qualitatively in line with the doping of phase 2. The good agreement between calculations and experiment supports our assignment of phase 1 to adsorbed Rb and phase 2 to intercalated Rb.

\section{Conclusions}
Applying various intercalation strategies for graphene often faces challenges such as the need of a high pressure of the gas containing the material to be intercalated and elevated temperatures might be required which can cause harmful reactions in the graphene lattice. In some cases smaller graphene islands are needed allowing the intercalant to pass through the edges\cite{Sutter:2010,granas:2012,Nilsson:2012,granas:2013}. It is essential to overcome these issues in order to tailor high quality graphene films with specific device characteristics. The results here demonstrate that it is possible to maintain the structural integrity of large area graphene using the intercalation approach, and moreover the system is stable up to a high temperature of 600~K.   

We find that using two intercalation steps a large span of carrier concentrations from strong $p$- to $n$-doping can be induced in graphene. This precise control of the doping level is in stark contrast to the situation in any bulk metal. This is exploited to vary the electron-phonon coupling and the Coulomb screening in graphene. Surprisingly, a finite screening is found in the presence of the metallic substrate contrary to what one might intuitively expect. These properties are central for the performance of devices that set out to exploit the Dirac particles for transport.   

The oxygen intercalated graphene system can be patterned with coexisting phases of adsorbed and intercalated rubidium that form $pn$-junctions within the carbon lattice with tunable hole concentration. This is possible in the first place because the system initially has a very large hole doping contrary to as-grown graphene on Ir(111). Systems like these could pave the way for very interesting relativistic quantum devices based on the phenomenon of Klein tunneling\cite{young:2009,stander:2009,Emtsev:2011}.

\section{Acknowledgements}
We gratefully acknowledge financial support from the European Research Council under ERC starting grant HPAH, No. 208344, the VILLUM foundation, The Danish Council for Independent Research / Technology and Production Sciences and the Lundbeck Foundation.

\providecommand*\mcitethebibliography{\thebibliography}
\csname @ifundefined\endcsname{endmcitethebibliography}
  {\let\endmcitethebibliography\endthebibliography}{}


\begin{mcitethebibliography}{51}
\providecommand*\natexlab[1]{#1}
\providecommand*\mciteSetBstSublistMode[1]{}
\providecommand*\mciteSetBstMaxWidthForm[2]{}
\providecommand*\mciteBstWouldAddEndPuncttrue
  {\def\EndOfBibitem{\unskip.}}
\providecommand*\mciteBstWouldAddEndPunctfalse
  {\let\EndOfBibitem\relax}
\providecommand*\mciteSetBstMidEndSepPunct[3]{}
\providecommand*\mciteSetBstSublistLabelBeginEnd[3]{}
\providecommand*\EndOfBibitem{}
\mciteSetBstSublistMode{f}
\mciteSetBstMaxWidthForm{subitem}{(\alph{mcitesubitemcount})}
\mciteSetBstSublistLabelBeginEnd
  {\mcitemaxwidthsubitemform\space}
  {\relax}
  {\relax}

\bibitem[Berger et~al.(2004)Berger, Song, Li, Li, Ogbazghi, Feng, Dai,
  Marchenkov, Conrad, First, and de~Heer]{Berger:2004}
Berger,~C.; Song,~Z.; Li,~T.; Li,~X.; Ogbazghi,~A.~Y.; Feng,~R.; Dai,~Z.;
  Marchenkov,~A.~N.; Conrad,~E.~H.; First,~P.~N. et~al.  Ultrathin Epitaxial
  Graphite: 2D Electron Gas Properties and a Route toward Graphene-based
  Nanoelectronics. \emph{The Journal of Physical Chemistry B} \textbf{2004},
  \emph{108}, 19912--19916\relax
\mciteBstWouldAddEndPuncttrue
\mciteSetBstMidEndSepPunct{\mcitedefaultmidpunct}
{\mcitedefaultendpunct}{\mcitedefaultseppunct}\relax
\EndOfBibitem
\bibitem[Emtsev et~al.(2009)Emtsev, Bostwick, Horn, Jobst, Kellogg, Ley,
  McChesney, Ohta, Reshanov, Rohrl, Rotenberg, Schmid, Waldmann, Weber, and
  Seyller]{Emtsev:2009}
Emtsev,~K.~V.; Bostwick,~A.; Horn,~K.; Jobst,~J.; Kellogg,~G.~L.; Ley,~L.;
  McChesney,~J.~L.; Ohta,~T.; Reshanov,~S.~A.; Rohrl,~J. et~al.  Towards
  wafer-size graphene layers by atmospheric pressure graphitization of silicon
  carbide. \emph{Nat Mater} \textbf{2009}, \emph{8}, 203--207\relax
\mciteBstWouldAddEndPuncttrue
\mciteSetBstMidEndSepPunct{\mcitedefaultmidpunct}
{\mcitedefaultendpunct}{\mcitedefaultseppunct}\relax
\EndOfBibitem
\bibitem[Sutter et~al.(2008)Sutter, Flege, and Sutter]{Sutter:2008}
Sutter,~P.~W.; Flege,~J.-I.; Sutter,~E.~A. Epitaxial graphene on ruthenium.
  \emph{Nat Mater} \textbf{2008}, \emph{7}, 406--411\relax
\mciteBstWouldAddEndPuncttrue
\mciteSetBstMidEndSepPunct{\mcitedefaultmidpunct}
{\mcitedefaultendpunct}{\mcitedefaultseppunct}\relax
\EndOfBibitem
\bibitem[Coraux et~al.(2008)Coraux, N'Diaye, Busse, and Michely]{Coraux:2008}
Coraux,~J.; N'Diaye,~A.~T.; Busse,~C.; Michely,~T. Structural coherency of
  graphene on Ir(111). \emph{Nano Lett.} \textbf{2008}, \emph{8}, 565--70\relax
\mciteBstWouldAddEndPuncttrue
\mciteSetBstMidEndSepPunct{\mcitedefaultmidpunct}
{\mcitedefaultendpunct}{\mcitedefaultseppunct}\relax
\EndOfBibitem
\bibitem[Li et~al.(2009)Li, Cai, An, Kim, Nah, Yang, Piner, Velamakanni, Jung,
  Tutuc, Banerjee, Colombo, and Ruoff]{Li:2009}
Li,~X.; Cai,~W.; An,~J.; Kim,~S.; Nah,~J.; Yang,~D.; Piner,~R.;
  Velamakanni,~A.; Jung,~I.; Tutuc,~E. et~al.  Large-Area Synthesis of
  High-Quality and Uniform Graphene Films on Copper Foils. \emph{Science}
  \textbf{2009}, \emph{324}, 1312--1314\relax
\mciteBstWouldAddEndPuncttrue
\mciteSetBstMidEndSepPunct{\mcitedefaultmidpunct}
{\mcitedefaultendpunct}{\mcitedefaultseppunct}\relax
\EndOfBibitem
\bibitem[Hao et~al.(2013)Hao, Bharathi, Wang, Liu, Chen, Nie, Wang, Chou, Tan,
  Fallahazad, Ramanarayan, Magnuson, Tutuc, Yakobson, {McCarty}, Zhang, Kim,
  Hone, Colombo, and Ruoff]{hao:2013}
Hao,~Y.; Bharathi,~M.~S.; Wang,~L.; Liu,~Y.; Chen,~H.; Nie,~S.; Wang,~X.;
  Chou,~H.; Tan,~C.; Fallahazad,~B. et~al.  The Role of Surface Oxygen in the
  Growth of Large Single-Crystal Graphene on Copper. \emph{Science}
  \textbf{2013}, \emph{342}, 720--723, {PMID:} 24158906\relax
\mciteBstWouldAddEndPuncttrue
\mciteSetBstMidEndSepPunct{\mcitedefaultmidpunct}
{\mcitedefaultendpunct}{\mcitedefaultseppunct}\relax
\EndOfBibitem
\bibitem[Novoselov et~al.(2012)Novoselov, Fal'ko, Colombo, Gellert, Schwab, and
  Kim]{Novoselov:2012}
Novoselov,~K.~S.; Fal'ko,~V.~I.; Colombo,~L.; Gellert,~P.~R.; Schwab,~M.~G.;
  Kim,~K. A roadmap for graphene. \emph{Nature} \textbf{2012}, \emph{490},
  192--200\relax
\mciteBstWouldAddEndPuncttrue
\mciteSetBstMidEndSepPunct{\mcitedefaultmidpunct}
{\mcitedefaultendpunct}{\mcitedefaultseppunct}\relax
\EndOfBibitem
\bibitem[Bostwick et~al.(2007)Bostwick, Ohta, Seyller, Horn, and
  Rotenberg]{Bostwick:2007}
Bostwick,~A.; Ohta,~T.; Seyller,~T.; Horn,~K.; Rotenberg,~E. Quasiparticle
  dynamics in graphene. \emph{Nature Physics} \textbf{2007}, \emph{3},
  36--40\relax
\mciteBstWouldAddEndPuncttrue
\mciteSetBstMidEndSepPunct{\mcitedefaultmidpunct}
{\mcitedefaultendpunct}{\mcitedefaultseppunct}\relax
\EndOfBibitem
\bibitem[Riedl et~al.(2009)Riedl, Coletti, Iwasaki, Zakharov, and
  Starke]{Riedl:2009}
Riedl,~C.; Coletti,~C.; Iwasaki,~T.; Zakharov,~A.~A.; Starke,~U.
  Quasi-Free-Standing Epitaxial Graphene on SiC Obtained by Hydrogen
  Intercalation. \emph{Physical Review Letters} \textbf{2009}, \emph{103},
  246804\relax
\mciteBstWouldAddEndPuncttrue
\mciteSetBstMidEndSepPunct{\mcitedefaultmidpunct}
{\mcitedefaultendpunct}{\mcitedefaultseppunct}\relax
\EndOfBibitem
\bibitem[Balog et~al.(2010)Balog, J{\o}rgensen, Nilsson, Andersen, Rienks,
  Bianchi, Fanetti, L{\ae}gsgaard, Baraldi, Lizzit, Sljivancanin, Besenbacher,
  Hammer, Pedersen, Hofmann, and Hornek{\ae}r]{Balog:2010}
Balog,~R.; J{\o}rgensen,~B.; Nilsson,~L.; Andersen,~M.; Rienks,~E.;
  Bianchi,~M.; Fanetti,~M.; L{\ae}gsgaard,~E.; Baraldi,~A.; Lizzit,~S. et~al.
  Band Gap Opening in Graphene Induced by Patterned Hydrogen Adsorption.
  \emph{Nature Materials} \textbf{2010}, \emph{9}, 315--319\relax
\mciteBstWouldAddEndPuncttrue
\mciteSetBstMidEndSepPunct{\mcitedefaultmidpunct}
{\mcitedefaultendpunct}{\mcitedefaultseppunct}\relax
\EndOfBibitem
\bibitem[Lizzit et~al.(2012)Lizzit, Larciprete, Lacovig, Dalmiglio, Orlando,
  Baraldi, Gammelgaard, Barreto, Bianchi, Perkins, and Hofmann]{Lizzit:2012}
Lizzit,~S.; Larciprete,~R.; Lacovig,~P.; Dalmiglio,~M.; Orlando,~F.;
  Baraldi,~A.; Gammelgaard,~L.; Barreto,~L.; Bianchi,~M.; Perkins,~E. et~al.
  Transfer-Free Electrical Insulation of Epitaxial Graphene from its Metal
  Substrate. \emph{Nano Letters} \textbf{2012}, \emph{12}, 4503--4507\relax
\mciteBstWouldAddEndPuncttrue
\mciteSetBstMidEndSepPunct{\mcitedefaultmidpunct}
{\mcitedefaultendpunct}{\mcitedefaultseppunct}\relax
\EndOfBibitem
\bibitem[Schumacher et~al.(2013)Schumacher, Wehling, Lazi\'c, Runte, F\"orster,
  Busse, Petrovi\'c, Kralj, Bl\"ugel, Atodiresei, Caciuc, and
  Michely]{schumacher:2013}
Schumacher,~S.; Wehling,~T.~O.; Lazi\'c,~P.; Runte,~S.; F\"orster,~D.~F.;
  Busse,~C.; Petrovi\'c,~M.; Kralj,~M.; Bl\"ugel,~S.; Atodiresei,~N. et~al.
  The Backside of Graphene: Manipulating Adsorption by Intercalation.
  \emph{Nano Letters} \textbf{2013}, \emph{13}, 5013--5019\relax
\mciteBstWouldAddEndPuncttrue
\mciteSetBstMidEndSepPunct{\mcitedefaultmidpunct}
{\mcitedefaultendpunct}{\mcitedefaultseppunct}\relax
\EndOfBibitem
\bibitem[Varykhalov et~al.(2008)Varykhalov, Sanchez-Barriga, Shikin, Biswas,
  Vescovo, Rybkin, Marchenko, and Rader]{Varykhalov:2008}
Varykhalov,~A.; Sanchez-Barriga,~J.; Shikin,~A.~M.; Biswas,~C.; Vescovo,~E.;
  Rybkin,~A.; Marchenko,~D.; Rader,~O. Electronic and Magnetic Properties of
  Quasifreestanding Graphene on Ni. \emph{Physical Review Letters}
  \textbf{2008}, \emph{101}, 157601\relax
\mciteBstWouldAddEndPuncttrue
\mciteSetBstMidEndSepPunct{\mcitedefaultmidpunct}
{\mcitedefaultendpunct}{\mcitedefaultseppunct}\relax
\EndOfBibitem
\bibitem[Sicot et~al.(2012)Sicot, Leicht, Zusan, Bouvron, Zander, Weser,
  Dedkov, Horn, and Fonin]{Sicot:2012}
Sicot,~M.; Leicht,~P.; Zusan,~A.; Bouvron,~S.; Zander,~O.; Weser,~M.;
  Dedkov,~Y.~S.; Horn,~K.; Fonin,~M. Size-Selected Epitaxial Nanoislands
  Underneath Graphene Moir{\`e} on Rh(111). \emph{ACS Nano} \textbf{2012},
  \emph{6}, 151--158\relax
\mciteBstWouldAddEndPuncttrue
\mciteSetBstMidEndSepPunct{\mcitedefaultmidpunct}
{\mcitedefaultendpunct}{\mcitedefaultseppunct}\relax
\EndOfBibitem
\bibitem[Marchenko et~al.(2012)Marchenko, Varykhalov, Scholz, Bihlmayer,
  Rashba, Rybkin, Shikin, and Rader]{marchenko:2012}
Marchenko,~D.; Varykhalov,~A.; Scholz,~M.~R.; Bihlmayer,~G.; Rashba,~E.~I.;
  Rybkin,~A.; Shikin,~A.~M.; Rader,~O. Giant Rashba splitting in graphene due
  to hybridization with gold. \emph{Nature Communications} \textbf{2012},
  \emph{3}, 1232\relax
\mciteBstWouldAddEndPuncttrue
\mciteSetBstMidEndSepPunct{\mcitedefaultmidpunct}
{\mcitedefaultendpunct}{\mcitedefaultseppunct}\relax
\EndOfBibitem
\bibitem[Emtsev et~al.(2008)Emtsev, Speck, Seyller, Ley, and
  Riley]{Emtsev:2008}
Emtsev,~K.~V.; Speck,~F.; Seyller,~T.; Ley,~L.; Riley,~J.~D. Interaction,
  growth, and ordering of epitaxial graphene on SiC(0001) surfaces: A
  comparative photoelectron spectroscopy study. \emph{Physical Review B}
  \textbf{2008}, \emph{77}, 155303\relax
\mciteBstWouldAddEndPuncttrue
\mciteSetBstMidEndSepPunct{\mcitedefaultmidpunct}
{\mcitedefaultendpunct}{\mcitedefaultseppunct}\relax
\EndOfBibitem
\bibitem[Bostwick et~al.(2010)Bostwick, Speck, Seyller, Horn, Polini, Asgari,
  MacDonald, and Rotenberg]{Bostwick:2010}
Bostwick,~A.; Speck,~F.; Seyller,~T.; Horn,~K.; Polini,~M.; Asgari,~R.;
  MacDonald,~A.~H.; Rotenberg,~E. Observation of Plasmarons in
  Quasi-Freestanding Doped Graphene. \emph{Science} \textbf{2010}, \emph{328},
  999--1002\relax
\mciteBstWouldAddEndPuncttrue
\mciteSetBstMidEndSepPunct{\mcitedefaultmidpunct}
{\mcitedefaultendpunct}{\mcitedefaultseppunct}\relax
\EndOfBibitem
\bibitem[Siegel et~al.(2011)Siegel, Park, Hwang, Deslippe, Fedorov, Louie, and
  Lanzara]{Siegel:2011}
Siegel,~D.~A.; Park,~C.-H.; Hwang,~C.; Deslippe,~J.; Fedorov,~A.~V.;
  Louie,~S.~G.; Lanzara,~A. Many-body interactions in quasi-freestanding
  graphene. \emph{Proceedings of the National Academy of Sciences}
  \textbf{2011}, \emph{108}, 11365--11369\relax
\mciteBstWouldAddEndPuncttrue
\mciteSetBstMidEndSepPunct{\mcitedefaultmidpunct}
{\mcitedefaultendpunct}{\mcitedefaultseppunct}\relax
\EndOfBibitem
\bibitem[Siegel et~al.(2012)Siegel, Hwang, Fedorov, and Lanzara]{Siegel:2012}
Siegel,~D.~A.; Hwang,~C.; Fedorov,~A.~V.; Lanzara,~A. Electron--phonon coupling
  and intrinsic bandgap in highly-screened graphene. \emph{New Journal of
  Physics} \textbf{2012}, \emph{14}, 095006\relax
\mciteBstWouldAddEndPuncttrue
\mciteSetBstMidEndSepPunct{\mcitedefaultmidpunct}
{\mcitedefaultendpunct}{\mcitedefaultseppunct}\relax
\EndOfBibitem
\bibitem[Walter et~al.(2011)Walter, Bostwick, Jeon, Speck, Ostler, Seyller,
  Moreschini, Chang, Polini, Asgari, MacDonald, Horn, and
  Rotenberg]{Walter:2011c}
Walter,~A.~L.; Bostwick,~A.; Jeon,~K.-J.; Speck,~F.; Ostler,~M.; Seyller,~T.;
  Moreschini,~L.; Chang,~Y.~J.; Polini,~M.; Asgari,~R. et~al.  Effective
  screening and the plasmaron bands in graphene. \emph{Phys. Rev. B}
  \textbf{2011}, \emph{84}, 085410\relax
\mciteBstWouldAddEndPuncttrue
\mciteSetBstMidEndSepPunct{\mcitedefaultmidpunct}
{\mcitedefaultendpunct}{\mcitedefaultseppunct}\relax
\EndOfBibitem
\bibitem[Larciprete et~al.(2012)Larciprete, Ulstrup, Lacovig, Dalmiglio,
  Bianchi, Mazzola, Hornek{\ae}r, Orlando, Baraldi, Hofmann, and
  Lizzit]{Larciprete:2012}
Larciprete,~R.; Ulstrup,~S.; Lacovig,~P.; Dalmiglio,~M.; Bianchi,~M.;
  Mazzola,~F.; Hornek{\ae}r,~L.; Orlando,~F.; Baraldi,~A.; Hofmann,~P. et~al.
  Oxygen Switching of the Epitaxial Graphene--Metal Interaction. \emph{ACS
  Nano} \textbf{2012}, \emph{6}, 9551--9558\relax
\mciteBstWouldAddEndPuncttrue
\mciteSetBstMidEndSepPunct{\mcitedefaultmidpunct}
{\mcitedefaultendpunct}{\mcitedefaultseppunct}\relax
\EndOfBibitem
\bibitem[Zhao et~al.(2011)Zhao, Kozlov, H\"{o}fert, Gotterbarm, Lorenz,
  Vi\~{n}es, Papp, G\"{o}rling, and Steinr\"{u}ck]{zhao:2011}
Zhao,~W.; Kozlov,~S.~M.; H\"{o}fert,~O.; Gotterbarm,~K.; Lorenz,~M. P.~A.;
  Vi\~{n}es,~F.; Papp,~C.; G\"{o}rling,~A.; Steinr\"{u}ck,~H.-P. Graphene on
  Ni(111): Coexistence of Different Surface Structures. \emph{The Journal of
  Physical Chemistry Letters} \textbf{2011}, \emph{2}, 759--764\relax
\mciteBstWouldAddEndPuncttrue
\mciteSetBstMidEndSepPunct{\mcitedefaultmidpunct}
{\mcitedefaultendpunct}{\mcitedefaultseppunct}\relax
\EndOfBibitem
\bibitem[Pletikosi\'{c} et~al.(2009)Pletikosi\'{c}, Kralj, Pervan, Brako,
  Coraux, N'Diaye, Busse, and Michely]{Pletikosic:2009}
Pletikosi\'{c},~I.; Kralj,~M.; Pervan,~P.; Brako,~R.; Coraux,~J.;
  N'Diaye,~A.~T.; Busse,~C.; Michely,~T. Dirac Cones and Minigaps for Graphene
  on Ir(111). \emph{Physical Review Letters} \textbf{2009}, \emph{102},
  056808\relax
\mciteBstWouldAddEndPuncttrue
\mciteSetBstMidEndSepPunct{\mcitedefaultmidpunct}
{\mcitedefaultendpunct}{\mcitedefaultseppunct}\relax
\EndOfBibitem
\bibitem[Kralj et~al.(2011)Kralj, Pletikosi\ifmmode~\acute{c}\else \'{c}\fi{},
  Petrovi\ifmmode~\acute{c}\else \'{c}\fi{}, Pervan, Milun, N'Diaye, Busse,
  Michely, Fujii, and Vobornik]{Kralj:2011}
Kralj,~M.; Pletikosi\ifmmode~\acute{c}\else \'{c}\fi{},~I.;
  Petrovi\ifmmode~\acute{c}\else \'{c}\fi{},~M.; Pervan,~P.; Milun,~M.;
  N'Diaye,~A.~T.; Busse,~C.; Michely,~T.; Fujii,~J.; Vobornik,~I. Graphene on
  Ir(111) characterized by angle-resolved photoemission. \emph{Phys. Rev. B}
  \textbf{2011}, \emph{84}, 075427\relax
\mciteBstWouldAddEndPuncttrue
\mciteSetBstMidEndSepPunct{\mcitedefaultmidpunct}
{\mcitedefaultendpunct}{\mcitedefaultseppunct}\relax
\EndOfBibitem
\bibitem[Starodub et~al.(2011)Starodub, Bostwick, Moreschini, Nie, Gabaly,
  McCarty, and Rotenberg]{Starodub:2011}
Starodub,~E.; Bostwick,~A.; Moreschini,~L.; Nie,~S.; Gabaly,~F.~E.;
  McCarty,~K.~F.; Rotenberg,~E. In-plane orientation effects on the electronic
  structure, stability, and Raman scattering of monolayer graphene on Ir(111).
  \emph{Phys. Rev. B} \textbf{2011}, \emph{83}, 125428\relax
\mciteBstWouldAddEndPuncttrue
\mciteSetBstMidEndSepPunct{\mcitedefaultmidpunct}
{\mcitedefaultendpunct}{\mcitedefaultseppunct}\relax
\EndOfBibitem
\bibitem[L{\ae}gsgaard et~al.(1988)L{\ae}gsgaard, Besenbacher, Mortensen, and
  Stensgaard]{laegsgaard:1988}
L{\ae}gsgaard,~E.; Besenbacher,~F.; Mortensen,~K.; Stensgaard,~I. A fully
  automated, thimble-sized scanning tunnelling microscope. \emph{Journal of
  Microscopy} \textbf{1988}, \emph{152}, 663--669\relax
\mciteBstWouldAddEndPuncttrue
\mciteSetBstMidEndSepPunct{\mcitedefaultmidpunct}
{\mcitedefaultendpunct}{\mcitedefaultseppunct}\relax
\EndOfBibitem
\bibitem[Hoffmann et~al.(2004)Hoffmann, S{\o}ndergaard, Schultz, Li, and
  Hofmann]{Hoffmann:2004}
Hoffmann,~S.~V.; S{\o}ndergaard,~C.; Schultz,~C.; Li,~Z.; Hofmann,~P. An
  undulator-based spherical grating monochromator beamline for angle-resolved
  photoemission spectroscopy. \emph{Nuclear Instruments and Methods in Physics
  Research Section A: Accelerators, Spectrometers, Detectors and Associated
  Equipment} \textbf{2004}, \emph{523}, 441--453\relax
\mciteBstWouldAddEndPuncttrue
\mciteSetBstMidEndSepPunct{\mcitedefaultmidpunct}
{\mcitedefaultendpunct}{\mcitedefaultseppunct}\relax
\EndOfBibitem
\bibitem[Enkovaara et~al.(2010)Enkovaara, Rostgaard, Mortensen, Chen, Du{\l}ak,
  Ferrighi, Gavnholt, Glinsvad, Haikola, Hansen, Kristoffersen, Kuisma, Larsen,
  Lehtovaara, Ljungberg, Lopez-Acevedo, Moses, Ojanen, Olsen, Petzold, Romero,
  Stausholm-M{\o}ller, Strange, Tritsaris, Vanin, Walter, Hammer, H\"{a}kkinen,
  Madsen, Nieminen, N{\o}rskov, Puska, Rantala, Schi{\o}tz, Thygesen, and
  Jacobsen]{Enkovaara2010}
Enkovaara,~J.; Rostgaard,~C.; Mortensen,~J.~J.; Chen,~J.; Du{\l}ak,~M.;
  Ferrighi,~L.; Gavnholt,~J.; Glinsvad,~C.; Haikola,~V.; Hansen,~H.~A. et~al.
  {IOPscience - Electronic structure calculations with GPAW: a real-space
  implementation of the projector augmented-wave method}. \emph{Journal of
  Physics: Condensed Matter} \textbf{2010}, \emph{22}, 253202\relax
\mciteBstWouldAddEndPuncttrue
\mciteSetBstMidEndSepPunct{\mcitedefaultmidpunct}
{\mcitedefaultendpunct}{\mcitedefaultseppunct}\relax
\EndOfBibitem
\bibitem[Bahn and Jacobsen(2002)Bahn, and Jacobsen]{Bahn2002}
Bahn,~S.~R.; Jacobsen,~K.~W. An object-oriented scripting interface to a legacy
  electronic structure code. \emph{Computing in Science and Engineering}
  \textbf{2002}, \emph{4}, 56--66\relax
\mciteBstWouldAddEndPuncttrue
\mciteSetBstMidEndSepPunct{\mcitedefaultmidpunct}
{\mcitedefaultendpunct}{\mcitedefaultseppunct}\relax
\EndOfBibitem
\bibitem[Klime\v{s} et~al.(2010)Klime\v{s}, Bowler, and
  Michaelides]{Klimes2010}
Klime\v{s},~J.; Bowler,~D.~R.; Michaelides,~A. Chemical accuracy for the van
  der Waals density functional. \emph{Journal of Physics: Condensed Matter}
  \textbf{2010}, \emph{22}, 022201\relax
\mciteBstWouldAddEndPuncttrue
\mciteSetBstMidEndSepPunct{\mcitedefaultmidpunct}
{\mcitedefaultendpunct}{\mcitedefaultseppunct}\relax
\EndOfBibitem
\bibitem[Mittendorfer et~al.(2011)]{Mittendorfer2011}
Mittendorfer,~F.; Garhofer,~A.; Redinger,~J.; Klime\v{s},~J.; Harl,~J.; Kresse,~G. Graphene on Ni(111): Strong interaction and weak adsorption. \emph{Phys. Rev. B}
  \textbf{2011}, \emph{84}, 201401(R)\relax
\mciteBstWouldAddEndPuncttrue
\mciteSetBstMidEndSepPunct{\mcitedefaultmidpunct}
{\mcitedefaultendpunct}{\mcitedefaultseppunct}\relax
\EndOfBibitem
\bibitem[Liu et~al.(2012)]{Liu2012}
Liu,~W.; Carrasco,~J.; Santra,~B.; Michalides,~A.; Scheffler,~M.; Tkatchenko,~A. Benzene adsorbed on metals: Concerted effect of covalency and van der Waals bonding. \emph{Phys. Rev. B}
  \textbf{2012}, \emph{86}, 245405\relax
\mciteBstWouldAddEndPuncttrue
\mciteSetBstMidEndSepPunct{\mcitedefaultmidpunct}
{\mcitedefaultendpunct}{\mcitedefaultseppunct}\relax
\EndOfBibitem
\bibitem[Petrovi\'c et~al.(2013)Petrovi\'c, \v{S}rut Raki\'c, Runte, Busse,
  Sadowski, Lazi\'c, Pletikosi\'c, Pan, Milun, Pervan, Atodiresei, Brako,
  \v{S}ok\v{c}evi\'c, Valla, Michely, and Kralj]{petrovic:2013}
Petrovi\'c,~M.; \v{S}rut Raki\'c,~I.; Runte,~S.; Busse,~C.; Sadowski,~J.~T.;
  Lazi\'c,~P.; Pletikosi\'c,~I.; Pan,~Z.-H.; Milun,~M.; Pervan,~P. et~al.  The
  mechanism of caesium intercalation of graphene. \emph{Nature Communications}
  \textbf{2013}, \emph{4}\relax
\mciteBstWouldAddEndPuncttrue
\mciteSetBstMidEndSepPunct{\mcitedefaultmidpunct}
{\mcitedefaultendpunct}{\mcitedefaultseppunct}\relax
\EndOfBibitem
\bibitem[Borca et~al.(2010)Borca, Barja, Garnica, Minniti, Politano,
  Rodriguez-Garcia, Hinarejos, Farias, Parga, and Miranda]{borca:2010}
Borca,~B.; Barja,~S.; Garnica,~M.; Minniti,~M.; Politano,~A.;
  Rodriguez-Garcia,~J.~M.; Hinarejos,~J.~J.; Farias,~D.; Parga,~A. L. V.~d.;
  Miranda,~R. Electronic and geometric corrugation of periodically rippled,
  self-nanostructured graphene epitaxially grown on Ru(0001). \emph{New Journal
  of Physics} \textbf{2010}, \emph{12}, 093018\relax
\mciteBstWouldAddEndPuncttrue
\mciteSetBstMidEndSepPunct{\mcitedefaultmidpunct}
{\mcitedefaultendpunct}{\mcitedefaultseppunct}\relax
\EndOfBibitem
\bibitem[N'Diaye et~al.(2008)N'Diaye, Coraux, Plasa, Busse, and
  Michely]{Diaye:2008}
N'Diaye,~A.~T.; Coraux,~J.; Plasa,~T.~N.; Busse,~C.; Michely,~T. Structure of
  epitaxial graphene on Ir(111). \emph{New Journal of Physics} \textbf{2008},
  \emph{10}, 043033\relax
\mciteBstWouldAddEndPuncttrue
\mciteSetBstMidEndSepPunct{\mcitedefaultmidpunct}
{\mcitedefaultendpunct}{\mcitedefaultseppunct}\relax
\EndOfBibitem
\bibitem[Krekelberg et~al.(2004)Krekelberg, Greeley, and
  Mavrikakis]{Krekelberg_JPCB_2004}
Krekelberg,~W.~P.; Greeley,~J.; Mavrikakis,~M. Atomic and Molecular Adsorption
  on Ir(111). \emph{The Journal of Physical Chemistry B} \textbf{2004},
  \emph{108}, 987--994\relax
\mciteBstWouldAddEndPuncttrue
\mciteSetBstMidEndSepPunct{\mcitedefaultmidpunct}
{\mcitedefaultendpunct}{\mcitedefaultseppunct}\relax
\EndOfBibitem
\bibitem[Shirley et~al.(1995)Shirley, Terminello, Santoni, and
  Himpsel]{Shirley:1995b}
Shirley,~E.~L.; Terminello,~L.~J.; Santoni,~A.; Himpsel,~F.~J.
  Brillouin-zone-selection effects in graphite photoelectron angular
  distributions. \emph{Phys. Rev. B} \textbf{1995}, \emph{51},
  13614--13622\relax
\mciteBstWouldAddEndPuncttrue
\mciteSetBstMidEndSepPunct{\mcitedefaultmidpunct}
{\mcitedefaultendpunct}{\mcitedefaultseppunct}\relax
\EndOfBibitem
\bibitem{mucha:2008}
M.~Mucha-Kruczy\'nski, O.~Tsyplyatyev, A.~Grishin, E.~{McCann}, Vladimir~I.
  Fal'ko, Aaron Bostwick, and Eli Rotenberg.
\newblock Characterization of graphene through anisotropy of constant-energy
  maps in angle-resolved photoemission.
\newblock {\em Physical Review B}, 2008, \textbf{77}, 195403\relax
\EndOfBibitem
\bibitem[Varykhalov et~al.(2012)Varykhalov, Marchenko, Scholz, Rienks, Kim,
  Bihlmayer, Sanchez-Barriga, and Rader]{varykhalov:2012r}
Varykhalov,~A.; Marchenko,~D.; Scholz,~M.~R.; Rienks,~E. D.~L.; Kim,~T.~K.;
  Bihlmayer,~G.; Sanchez-Barriga,~J.; Rader,~O. Ir(111) Surface State with
  Giant Rashba Splitting Persists under Graphene in Air. \emph{Physical Review
  Letters} \textbf{2012}, \emph{108}, 066804\relax
\mciteBstWouldAddEndPuncttrue
\mciteSetBstMidEndSepPunct{\mcitedefaultmidpunct}
{\mcitedefaultendpunct}{\mcitedefaultseppunct}\relax
\EndOfBibitem
\bibitem[Hofmann et~al.(2009)Hofmann, Sklyadneva, Rienks, and
  Chulkov]{Hofmann:2009b}
Hofmann,~P.; Sklyadneva,~I.~Y.; Rienks,~E. D.~L.; Chulkov,~E.~V.
  Electron--phonon coupling at surfaces and interfaces. \emph{New Journal of
  Physics} \textbf{2009}, \emph{11}, 125005\relax
\mciteBstWouldAddEndPuncttrue
\mciteSetBstMidEndSepPunct{\mcitedefaultmidpunct}
{\mcitedefaultendpunct}{\mcitedefaultseppunct}\relax
\EndOfBibitem
\bibitem[Das~Sarma et~al.(2007)Das~Sarma, Hwang, and Tse]{Das-Sarma:2007}
Das~Sarma,~S.; Hwang,~E.~H.; Tse,~W.-K. Many-body interaction effects in doped
  and undoped graphene: Fermi liquid versus non-Fermi liquid. \emph{Phys. Rev.
  B} \textbf{2007}, \emph{75}, 121406\relax
\mciteBstWouldAddEndPuncttrue
\mciteSetBstMidEndSepPunct{\mcitedefaultmidpunct}
{\mcitedefaultendpunct}{\mcitedefaultseppunct}\relax
\EndOfBibitem
\bibitem[Das~Sarma and Hwang(2013)Das~Sarma, and Hwang]{DasSarma:2013}
Das~Sarma,~S.; Hwang,~E.~H. Velocity renormalization and anomalous
  quasiparticle dispersion in extrinsic graphene. \emph{Physical Review B}
  \textbf{2013}, \emph{87}, 045425\relax
\mciteBstWouldAddEndPuncttrue
\mciteSetBstMidEndSepPunct{\mcitedefaultmidpunct}
{\mcitedefaultendpunct}{\mcitedefaultseppunct}\relax
\EndOfBibitem
\bibitem[Pletikosi\ifmmode~\acute{c}\else \'{c}\fi{}
  et~al.(2012)Pletikosi\ifmmode~\acute{c}\else \'{c}\fi{}, Kralj, Milun, and
  Pervan]{Pletikosic:2012}
Pletikosi\ifmmode~\acute{c}\else \'{c}\fi{},~I.; Kralj,~M.; Milun,~M.;
  Pervan,~P. Finding the bare band: Electron coupling to two phonon modes in
  potassium-doped graphene on Ir(111). \emph{Phys. Rev. B} \textbf{2012},
  \emph{85}, 155447\relax
\mciteBstWouldAddEndPuncttrue
\mciteSetBstMidEndSepPunct{\mcitedefaultmidpunct}
{\mcitedefaultendpunct}{\mcitedefaultseppunct}\relax
\EndOfBibitem
\bibitem[Johannsen et~al.(2013)Johannsen, Ulstrup, Bianchi, Hatch, Guan,
  Mazzola, Hornek{\ae}r, Fromm, Raidel, Seyller, and Hofmann]{Johannsen:2013}
Johannsen,~J.~C.; Ulstrup,~S.; Bianchi,~M.; Hatch,~R.; Guan,~D.; Mazzola,~F.;
  Hornek{\ae}r,~L.; Fromm,~F.; Raidel,~C.; Seyller,~T. et~al.  Electron-phonon
  coupling in quasi-free-standing graphene. \emph{Journal of Physics: Condensed
  Matter} \textbf{2013}, \emph{25}, 094001\relax
\mciteBstWouldAddEndPuncttrue
\mciteSetBstMidEndSepPunct{\mcitedefaultmidpunct}
{\mcitedefaultendpunct}{\mcitedefaultseppunct}\relax
\EndOfBibitem
\bibitem[Rytk\'onen et~al.(2007)Rytk\'onen, Akola, and
  Manninen]{Rytkonen_PRB_2007}
Rytk\'onen,~K.; Akola,~J.; Manninen,~M. Density functional study of
  alkali-metal atoms and monolayers on graphite (0001). \emph{Phys. Rev. B}
  \textbf{2007}, \emph{75}, 075401\relax
\mciteBstWouldAddEndPuncttrue
\mciteSetBstMidEndSepPunct{\mcitedefaultmidpunct}
{\mcitedefaultendpunct}{\mcitedefaultseppunct}\relax
\EndOfBibitem
\bibitem[Vilhelmsen and Hammer(2012)Vilhelmsen, and
  Hammer]{Vilhelmsen_PRL_2012}
Vilhelmsen,~L.~B.; Hammer,~B. Systematic Study of Au$_6$ to Au$_{12}$ Gold
  Clusters on MgO(100) F Centers Using Density-Functional Theory. \emph{Phys.
  Rev. Lett.} \textbf{2012}, \emph{108}, 126101\relax
\mciteBstWouldAddEndPuncttrue
\mciteSetBstMidEndSepPunct{\mcitedefaultmidpunct}
{\mcitedefaultendpunct}{\mcitedefaultseppunct}\relax
\EndOfBibitem
\bibitem[Sutter et~al.(2010)Sutter, Sadowski, and Sutter]{Sutter:2010}
Sutter,~P.; Sadowski,~J.~T.; Sutter,~E.~A. Chemistry under Cover: Tuning
  Metal-Graphene Interaction by Reactive Intercalation. \emph{Journal of the
  American Chemical Society} \textbf{2010}, \emph{132}, 8175--8179\relax
\mciteBstWouldAddEndPuncttrue
\mciteSetBstMidEndSepPunct{\mcitedefaultmidpunct}
{\mcitedefaultendpunct}{\mcitedefaultseppunct}\relax
\EndOfBibitem
\bibitem[Gr{\aa}n\"{a}s et~al.(2012)Gr{\aa}n\"{a}s, Knudsen, Schr{\o}der,
  Gerber, Busse, Arman, Schulte, Andersen, and Michely]{granas:2012}
Gr{\aa}n\"{a}s,~E.; Knudsen,~J.; Schr{\o}der,~U.~A.; Gerber,~T.; Busse,~C.;
  Arman,~M.~A.; Schulte,~K.; Andersen,~J.~N.; Michely,~T. Oxygen Intercalation
  under Graphene on Ir(111): Energetics, Kinetics, and the Role of Graphene
  Edges. \emph{{ACS} Nano} \textbf{2012}, \emph{6}, 9951--9963\relax
\mciteBstWouldAddEndPuncttrue
\mciteSetBstMidEndSepPunct{\mcitedefaultmidpunct}
{\mcitedefaultendpunct}{\mcitedefaultseppunct}\relax
\EndOfBibitem
\bibitem[Nilsson et~al.(2012)Nilsson, Andersen, Balog, L{\ae}gsgaard, Hofmann,
  Besenbacher, Hammer, Stensgaard, and Hornek{\ae}r]{Nilsson:2012}
Nilsson,~L.; Andersen,~M.; Balog,~R.; L{\ae}gsgaard,~E.; Hofmann,~P.;
  Besenbacher,~F.; Hammer,~B.; Stensgaard,~I.; Hornek{\ae}r,~L. Graphene
  Coatings: Probing the Limits of the One Atom Thick Protection Layer.
  \emph{ACS Nano} \textbf{2012}, \emph{6}, 10258--10266\relax
\mciteBstWouldAddEndPuncttrue
\mciteSetBstMidEndSepPunct{\mcitedefaultmidpunct}
{\mcitedefaultendpunct}{\mcitedefaultseppunct}\relax
\EndOfBibitem
\bibitem[Gr{\aa}n\"{a}s et~al.(2013)Gr{\aa}n\"{a}s, Andersen, Arman, Gerber,
  Hammer, Schnadt, Andersen, Michely, and Knudsen]{granas:2013}
Gr{\aa}n\"{a}s,~E.; Andersen,~M.; Arman,~M.~A.; Gerber,~T.; Hammer,~B.;
  Schnadt,~J.; Andersen,~J.~N.; Michely,~T.; Knudsen,~J. {CO} Intercalation of
  Graphene on Ir(111) in the Millibar Regime. \emph{The Journal of Physical
  Chemistry C} \textbf{2013}, \emph{117}, 16438--16447\relax
\mciteBstWouldAddEndPuncttrue
\mciteSetBstMidEndSepPunct{\mcitedefaultmidpunct}
{\mcitedefaultendpunct}{\mcitedefaultseppunct}\relax
\EndOfBibitem
\bibitem[Young and Kim(2009)Young, and Kim]{young:2009}
Young,~A.~F.; Kim,~P. Quantum interference and Klein tunnelling in graphene
  heterojunctions. \emph{Nature Physics} \textbf{2009}, \emph{5},
  222--226\relax
\mciteBstWouldAddEndPuncttrue
\mciteSetBstMidEndSepPunct{\mcitedefaultmidpunct}
{\mcitedefaultendpunct}{\mcitedefaultseppunct}\relax
\EndOfBibitem
\bibitem[Stander et~al.(2009)Stander, Huard, and
  Goldhaber-Gordon]{stander:2009}
Stander,~N.; Huard,~B.; Goldhaber-Gordon,~D. Evidence for Klein Tunneling in
  Graphene p-n Junctions. \emph{Physical Review Letters} \textbf{2009},
  \emph{102}, 026807\relax
\mciteBstWouldAddEndPuncttrue
\mciteSetBstMidEndSepPunct{\mcitedefaultmidpunct}
{\mcitedefaultendpunct}{\mcitedefaultseppunct}\relax
\EndOfBibitem
\bibitem[Emtsev et~al.(2011)Emtsev, Zakharov, Coletti, Forti, and
  Starke]{Emtsev:2011}
Emtsev,~K.~V.; Zakharov,~A.~A.; Coletti,~C.; Forti,~S.; Starke,~U. Ambipolar
  doping in quasifree epitaxial graphene on SiC(0001) controlled by Ge
  intercalation. \emph{Phys. Rev. B} \textbf{2011}, \emph{84}, 125423\relax
\mciteBstWouldAddEndPuncttrue
\mciteSetBstMidEndSepPunct{\mcitedefaultmidpunct}
{\mcitedefaultendpunct}{\mcitedefaultseppunct}\relax
\EndOfBibitem
\end{mcitethebibliography}
\end{document}